  \documentclass[structabstract]{aa}
\usepackage{graphicx}
\usepackage{txfonts}
\usepackage{natbib}

\bibpunct[]{(}{)}{;}{a}{}{,}

\newcommand{\degC}{~$^{\circ}$C }

\begin{document}
   \title{The quest for complex molecules in space: Laboratory spectroscopy of 
          $n$-butyl cyanide, $n$-C$_4$H$_9$CN, in the millimeter wave region and its 
          astronomical search in \object{Sagittarius~B2(N)}}

   \author{M.~H. Ordu\inst{1}
           \and
           H.~S.~P. M{\"u}ller\inst{1,2}
           \and
           A. Walters\inst{3,4}
           \and
           M. Nu\~{n}ez\inst{3,4}
           \and
           F. Lewen\inst{1}
           \and
           A. Belloche\inst{2}
           \and
           K.~M. Menten\inst{2}
           \and
           S. Schlemmer\inst{1}}

   \institute{I.~Physikalisches Institut, Universit{\"a}t zu K{\"o}ln,
              Z{\"u}lpicher Str. 77, 50937 K{\"o}ln, Germany\\
              \email{hspm@ph1.uni-koeln.de}
         \and
              Max-Planck Institut f\"ur Radioastronomie, Auf dem H\"ugel 69, 
              53121 Bonn, Germany
         \and
              Universit\'e de Toulouse, UPS-OMP, IRAP, 
              Toulouse, France
         \and
              CNRS, IRAP, 
              9~Av. colonel Roche, BP~44346, 
              31028 Toulouse cedex~4, France}

   \date{Received 26 December 2011 / Accepted 9 March 2012}

  \abstract
{The saturated $n$-propyl cyanide was recently detected in Sagittarius~B2(N). 
The next larger unbranched alkyl cyanide is $n$-butyl cyanide.}
{We provide accurate rest frequency predictions beyond the millimeter wave 
range to search for this molecule in the Galactic center source Sagittarius~B2(N)
and facilitate its detection in space.}
{We investigated the laboratory rotational spectrum of $n$-butyl cyanide 
between 75~GHz and 348~GHz. We searched for emission lines produced by the 
molecule in our sensitive IRAM~30\,m molecular line survey of Sagittarius~B2(N).}
{We identified more than one thousand rotational transitions in the laboratory 
for each of the three conformers for which limited data had been obtained previously 
in a molecular beam microwave study. The quantum number range was greatly 
extended to $J \approx 120$ or more and $K_a > 35$, resulting in accurate 
spectroscopic parameters and accurate rest frequency calculations up to about 
500~GHz for strong to moderately weak transitions of the two lower energy 
conformers. 
Upper limits to the column densities of $N \le 3 \times 10^{15}$~cm$^{-2}$ 
and $8 \times 10^{15}$~cm$^{-2}$ were derived towards Sagittarius~B2(N) for the 
two lower energy conformers, $anti$-$anti$ and $gauche$-$anti$, respectively.}
{Our present data will be helpful for identifying $n$-butyl cyanide 
at millimeter or longer wavelengths with radio telescope arrays 
such as ALMA, NOEMA, or EVLA. In particular, its detection in Sagittarius~B2(N) 
with ALMA seems feasible.}

\titlerunning{Millimeter wave spectroscopy of $n$-butyl cyanide}

\maketitle
\hyphenation{For-schungs-ge-mein-schaft}

%

\section{Introduction}
\label{intro}

Cyanides account for 20 of the approximately 165 molecules detected in 
the interstellar medium or in the circumstellar envelopes of late-type 
stars\footnote{See e.\,g. the Molecules in Space page, 
http://www.astro.uni-koeln.de/cdms/molecules, of the Cologne Database 
for Molecular Spectroscopy, CDMS}; not included in this number are 
isocyanides and other molecules containing the CN moiety. 
Saturated cyanides are usually found in high-mass star-forming regions, 
as are other saturated molecules. One part of the Galactic center 
source Sagittarius (Sgr for short) B2(N) was nicknamed Large Molecule Heimat 
because of the many and in part rather complex molecules found there, 
many of them for the first time \citep{name_LMH,KMM-2004}. 
Line surveys of the two hot cores Sgr~B2(N) and Sgr~B2(M) at 3~mm, 
with selected observations at higher frequencies, have been carried out 
with the IRAM 30~m telescope to investigate the molecular complexity 
in these prolific sources. 
Two cyanides, aminoacetonitrile \citep[NH$_2$CH$_2$CN,][]{det_AAN}, a 
potential precursor to glycine, and $n$-propyl cyanide along with ethyl formate 
\citep[$n$-C$_3$H$_7$CN and C$_2$H$_5$OCHO,][]{det-PrCN_EtFo} were detected 
for the first time in space toward Sgr~B2(N) in the course of this study. 
In addition, $^{13}$C isotopologs of vinyl cyanide (C$_2$H$_3$CN) were 
detected for the first time in the interstellar medium, and those 
of ethyl cyanide (C$_2$H$_5$CN) for the first time in this source 
\citep{13C-VyCN_2008}; the latter had been identified shortly before 
in a line survey of Orion~KL \citep{13C-EtCN_2007}.

With the recent detection of $n$-propyl cyanide, the series of unbranched 
alkyl cyanides detected in the interstellar medium now contains 
three members: methyl cyanide, ethyl cyanide, and $n$-propyl cyanide. 
The chemical models of \citet{det-PrCN_EtFo} succeeded in reproducing 
the measured column density ratios with a sequential, piecewise construction 
of these alkyl cyanides from their constituent functional groups on 
the grain surfaces, which suggests that this chemical process is their 
most likely formation route. In addition, these models provide valuable 
constraints on the possible chemical pathways leading to the formation 
of complex organic molecules. Detecting the next member in the series 
of unbranched alkyl cyanides will indeed be a step forward in our 
understanding of the degree of chemical complexity that can be reached 
in the interstellar medium.


\begin{figure*}
\centering
\includegraphics[angle=0,width=17cm]{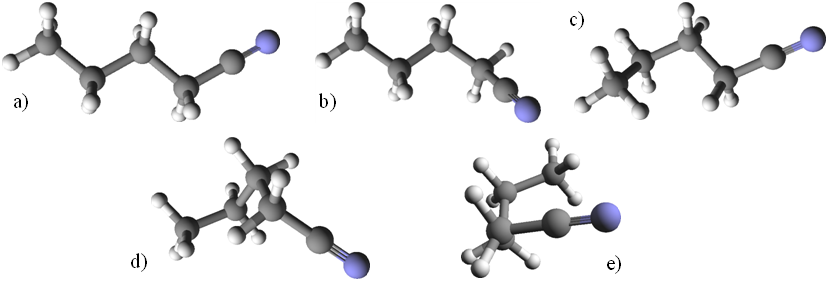}
  \caption{The conformers of $n$-butyl cyanide. Those investigated in 
  the present work are given in the upper row: a) $anti$-$anti$, 
  b) $gauche$-$anti$, and c) $anti$-$gauche$; the others are in the lower 
  row: d) $gauche$-$gauche$' and e) $gauche$-$gauche$. The ordering is 
  from a) being lowest to e) being highest in energy. 
  The C and N atoms are indicated by gray and violet ''spheres'', 
  respectively, and the H atoms by small, light gray ''spheres''.}
\label{conformers}
\end{figure*}

A complication of especially high-resolution spectroscopic investigations 
of $n$-butyl cyanide arises from the different conformations in which the 
molecule may occur. The $anti$-$anti$ (or $AA$) conformer possesses a planar 
zig-zag structure of the heavy atoms as demonstrated in Fig.~\ref{conformers}. 
Rotating either the cyano group or the terminal methyl group by about 
120$^{\rm o}$ yields the $gauche$-$anti$ ($GA$) and $anti$-$gauche$ ($AG$) 
conformer, respectively. Since either rotation can be performed clockwise 
as well as counterclockwise, both conformers are doubly degenerate 
with respect to the $AA$ conformer. Rotating both groups yields 
two different conformers because they can be rotated to either the same 
or the opposite side; the former is called $gauche$-$gauche$ ($GG$), 
while the latter is called $gauche$-$gauche$' ($GG$'). Both conformers 
are also doubly degenerate with respect to the $AA$ conformer. 

\citet{n-BuCN_IR_1989} studied the infrared spectrum of $n$-butyl cyanide 
in the gas phase. He stated that there are four spectroscopically 
distinguishable conformers. Among these, $GA$ was found to be the most abundant 
one (46\,\%), $AA$ was slightly less abundant (30\,\%), and even $AG$ 
(13\,\%) and a fourth one, presumably rather $GG$' than $GG$ (11\,\%), 
were present. Molecular mechanics\footnote{Molecular mechanics is a 
theoretical model usually based on simplified molecular force fields 
describing in particular larger molecules.} calculations yielded $AA$ 
as being the lowest in energy with $GA$, $AG$, and $GG$' being higher 
by 0.54, 3.68, and 4.31~kJ/mol (or 65, 443, and 519~K), respectively, 
roughly in accordance with the thermal population of the conformers. 
Very similar results were obtained for the isoelectronic 1-hexyne. 

\citet{n-BuCN_rot_1997} studied the rotational spectrum of $n$-butyl 
cyanide between 5~GHz and 22~GHz employing molecular beam Fourier 
transform microwave (MB-FTMW) spectroscopy. Three conformers were 
identified, $AA$, $GA$, and $AG$. The first two were estimated to 
be lower in energy than the latter because of the intensities. 
These findings are in accordance with those from gas-phase 
infrared spectroscopy mentioned in the previous paragraph. 
More than 20 rotational transitions were recorded for each conformer 
with $J$ and $K_a$ extending to 8 and 2, respectively. $^{14}$N 
quadrupole splitting was observed for some transitions, and 
the diagonal components of the quadrupole tensor were determined. 

Quantum chemical investigations into the energetics of the $n$-butyl 
cyanide conformers have been performed to our knowledge only by 
\citet{n-BuCN_ai_2002}. They carried out Hartree-Fock (HF) and second 
order M{\o}ller-Plesset (MP2) ab-initio calculations for all five 
conformers of $n$-butyl cyanide. The HF calculations yielded $AA$ and 
$GA$ as low-energy conformers with $AG$ and $GG$' higher by about 
4~kJ/mol ($\sim$500~K) and $GG$ much higher by more than 8~kJ/mol 
($\sim$1000~K). This energy ordering is compatible with both 
spectroscopic studies and is adopted in the present investigation. 
The higher level MP2 calculations lowered the relative energy of all 
conformers with a $gauche$ orientation of the CN group, but raised 
the relative energy of $AA$. As a consequence, $AA$ and $GG$' were 
calculated to be essentially isoenergetic. These energies, however, 
are incompatible with experimental results and are thus 
discarded.

The previously obtained spectroscopic parameters for three conformers 
of $n$-butyl cyanide \citep{n-BuCN_rot_1997} permit reliable predictions 
up to 40~GHz, partially even 60~GHz. However, already in the 3-mm region, 
deviations of a few megahertz must be expected. Therefore, we decided 
to investigate the rotational spectrum of this alkyl cyanide at shorter 
millimeter wavelengths to facilitate radio-astronomical observations 
throughout this region. The list of experimental transition frequencies 
was very greatly increased in the course of the present investigation, 
as were the frequency and quantum number ranges, permitting strong and 
medium strong transitions to be predicted accurately into the lower 
submillimeter-wave region. In addition, we used the data to search 
for the two lowest energy conformers of $n$-butyl cyanide in our 3~mm 
line survey of Sgr~B2(N).

\section{Experimental details}
\label{exptl}

The spectral ranges 75$-$131~GHz and 200$-$223~GHz were studied 
with a new, fully solid-state, source-modulated spectrometer. 
Detection of the intrinsically weak rotational transitions exhibited 
by complex molecules required the integration of an absorption cell 
with an effective path length of 44 meters. This double-pass cell 
comprises three Pyrex tubes of 10~cm diameter and almost equal lengths, 
and employs both a rooftop mirror, which rotates the polarization 
by 90 degrees, and a polarizing grid to separate the orthogonal source 
and detector beams. Millimeter-wave spectrometers with similar 
double-pass optics have been previously described in the literature, 
e.\,g. by \citet{O3-broadening-JPL}.

Observations of butyl cyanide absorption features were accomplished 
by establishing a slow gas flow through the absorption cell. This 
practice allowed the maintenance of the low and constant pressure 
(around $6 \times 10^{-3}$~mbar) required to avoid pressure-broadening 
over long data acquisition times. Heating of the gas inlet valve to 
about 60\degC was also found to be necessary to avoid sample condensation.

The spectrometer's radiation source consists of a microwave synthesizer 
covering the 10 to 43.5~GHz range whose output has been amplified, 
followed by a 3-mm band doubler and a chain of three 2-mm band doublers, 
respectively. The detector is a waveguide-mounted GaAs Schottky diode, 
optimally biased with a custom current source constructed at the 
Universit\"at zu K\"oln. The detector output is processed by a 
lock-in amplifier with a time constant of 50~ms; digital averaging 
of six to eight points typically results in an effective time constant 
of 300$-$400~ms. 

Additional measurements between 301 and 302 as well as between 342~GHz 
and 348~GHz employed a phase-locked backward-wave oscillator (BWO) 
spectrometer, the operating principles of which are described in greater 
detail in \citet{Winnewisser_BWO_solo} and \citet{Lewen_BWO_pulsed}. 
In the current spectrometer setup, phase locking is achieved by means 
of a harmonic mixer producing the third harmonic of a frequency-tripled 
microwave signal. The BWO beam thus stabilized in frequency is directed 
through a 3.4~m long, 10~cm diameter Pyrex cell and detected with a 
helium-cooled InSb hot-electron bolometer. 
Double-passing or larger cell dimensions are not needed here because 
of the high output power of the BWO (30~mW) and the low noise level 
of the bolometer. Frequency modulation and referencing are performed 
as described above.

\section{Results}

\subsection{Laboratory spectroscopy}
\label{obs}

The rotational spectrum of a large molecule is denser than that of a 
smaller molecule because the absorbed or emitted flux is distributed 
over more transitions spanning a larger range of rotational quantum 
numbers $J$ and $K_a$, yielding larger partition function values, 
and possibly over several conformers, increasing partition function 
values even further. In addition, transitions are observed not only 
in the ground vibrational state, but also in an increasing number of 
excited states for increasingly heavier molecules. Some examples 
illustrate the increase in line density with increasing molecular 
complexity. The rotational spectra of, e.\,g., CH$_3$CN and SO$_2$ 
are comparatively sparse such that transitions of $^{13}$CH$_3 ^{13}$CN 
\citep{CH3CN_rot_2009} and SO$^{17}$O \citep{SOO-17_rot_2000}, respectively, 
can be studied rather extensively by conventional absorption spectroscopy 
in a sample of natural isotopic composition. And while $^{13}$C, 
and even $^{15}$N, isotopologs of vinyl cyanide of natural isotopic 
composition were studied \citep{13C-VyCN_2008,VyCN_rot_2009}, 
the $^{13}$C isotopologs of the only slightly larger (2 H atoms) 
ethyl cyanide molecule were studied in isotopically enriched samples 
\citep{13C-EtCN_2007} because the increase in the number of rotational 
states in each vibrational state as well as the increase in the number 
of low-lying vibrational states result in a spectrum containing many 
more lines.

The rotational spectrum of $n$-butyl cyanide was studied in the 
microwave region initially by absorption spectroscopy by 
\citet{n-BuCN_rot_1997}, but their spectrum turned out to be 
too dense to perform an in-depth analysis. These authors hence used 
MB-FTMW, which is a very good alternative way of studying the 
rotational spectrum of a large complex molecule, in particular for 
initial assignments, because low rotational temperatures, as low as 
1~K, permit only a very small number of rotational levels to be 
populated compared to room temperature. \citet{n-BuCN_rot_1997} 
identified more than 20 transitions each between 5~GHz and 22~GHz 
for the $AA$, $GA$, and $AG$ conformers with quantum numbers $J$ 
and $K_a$ up to 8 and 2, respectively. All three conformers are 
asymmetric tops where Ray's asymmetry parameter $\kappa = 
(2B - A - C)/(A - C)$ is close to the prolate limit of $-1$. 
The values for the $AA$, $GA$, and $AG$ conformers are 
$-0.9898$, $-0.9229$, and $-0.9866$, respectively. 
The dipole moment components for each conformer were estimated 
from that of CH$_3$CN by rotating the latter molecule such that 
the CN bonds are parallel. The authors obtained $\mu _a = 3.4$~D 
and $\mu _b = 1.9$~D for the $AA$ conformer, and $\mu _c = 0$ 
because of the symmetry. The values for the $GA$ and $AG$ 
conformers were $\mu _a = 2.3$~D, $\mu _b = 3.1$~D, and 
$\mu _c \approx 0$~D and $\mu _a = 3.7$~D, $\mu _b \approx 0$~D, 
and $\mu _c = 1.2$~D, respectively. The diagonal elements of 
the $^{14}$N quadrupole tensor were estimated equivalently 
from CH$_3$CN data.


\begin{figure}
\includegraphics[angle=0,width=9cm]{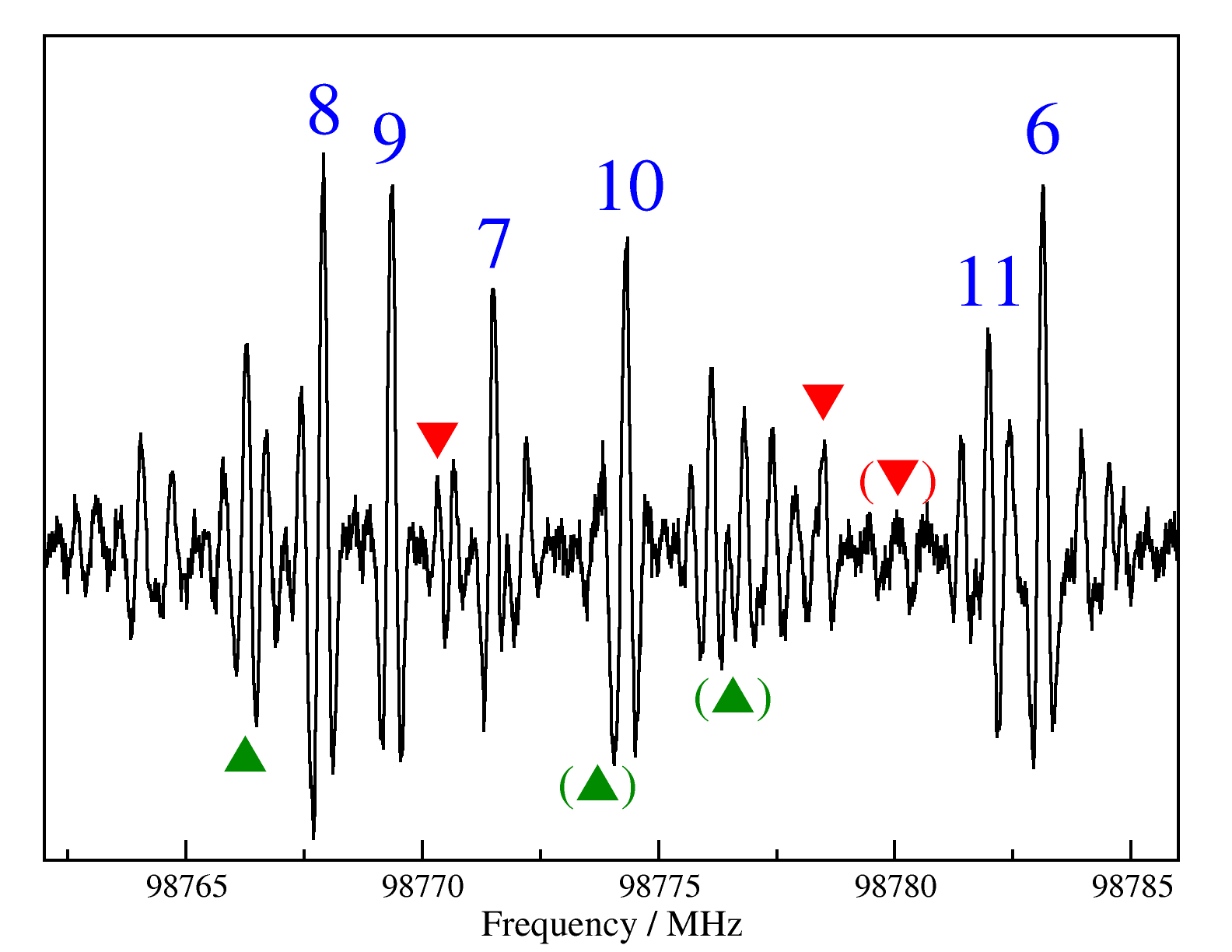}
  \caption{Section of the millimeter-wave spectrum of $n$-C$_4$H$_9$CN 
   displaying features caused by all three conformers investigated in the 
   present analysis. The $K' = K''$ quantum numbers for the $J = 38 - 37$ 
   transitions of the $AA$ conformer are given in blue (the $K_c$ quantum 
   numbers have been omitted because the asymmetry splitting was not resolved 
   for these transitions). Transitions of the $GA$ and $AG$ conformers 
   have been marked with triangles pointing upward (green) and downward (red), 
   respectively. Triangles in parentheses are used for overlapped lines or 
   lines with bad line shape. The $AG$ lines have $J = 34 - 33$, and
   $K_a = 27$, 4, and 28 from left to right; $K_c = J - K_a + 1$ for the 
   $K_a = 4$ line. The $GA$ lines belong to $b$-type $Q$-branch transitions 
   with quantum numbers $J_{K_a,K_c} = 40_{9,31} - 40_{8,32}$, 
   $69_{7,62} - 69_{6,63}$, and $95_{11,84} - 95_{10,85}$ from left to right.
   The lines are presented as second derivatives of a Gaussian 
   line shape because of the 2$f$-modulation.}
\label{lab-spectrum}
\end{figure}



\begin{table*}
\begin{center}
\caption{Spectroscopic parameters$^a$ (MHz) of three $n$-butyl cyanide conformers.}
\label{spec-parameter}
\renewcommand{\arraystretch}{1.10}
\begin{tabular}[t]{lr@{}lr@{}lr@{}l}
\hline \hline
Parameter & \multicolumn{2}{c}{$AA$} & \multicolumn{2}{c}{$AG$} & \multicolumn{2}{c}{$GA$} \\
\hline
$A$                      & 15\,028&.687\,18(46)   & 11\,887&.576\,88(215)  & 7\,635&.624\,96(22)   \\
$B$                      &  1\,334&.106\,343(23)  &  1\,486&.185\,826(80)  & 1\,788&.633\,540(31)  \\
$C$                      &  1\,263&.857\,661(22)  &  1\,415&.762\,264(76)  & 1\,554&.221\,920(33)  \\
$D_K \times 10^3$        &     214&.378\,0(86)    &     111&.73(207)       &     52&.400\,8(12)    \\
$D_{JK} \times 10^3$     &    $-$7&.543\,09(14)   &    $-$2&.733\,31(18)   &   $-$9&.899\,90(15)   \\
$D_J \times 10^6$        &     145&.218\,0(29)    &     227&.032\,0(80)    &    861&.210\,5(67)    \\
$d_1 \times 10^6$        &   $-$22&.019\,6(20)    &   $-$15&.742\,8(83)    & $-$233&.352\,6(72)    \\
$d_2 \times 10^6$        &    $-$0&.304\,89(26)   &    $-$2&.284\,0(59)    &   $-$8&.493\,0(35)    \\
$H_K \times 10^6$        &       4&.771(32)       &       2&.0$^b$         &      1&.452\,8(26)    \\
$H_{KJ} \times 10^9$     &   $-$53&.11(28)        &      13&.05(33)        & $-$351&.59(37)        \\
$H_{JK} \times 10^9$     &    $-$6&.427(11)       &    $-$5&.849(8)        &     14&.236(19)       \\
$H_J \times 10^{12}$     &      98&.54(13)        &     137&.00(33)        &    551&.63(47)        \\
$h_1 \times 10^{12}$     &      31&.37(15)        &   $-$30&.46(44)        &    263&.99(156)       \\
$h_2 \times 10^{12}$     &        &               &        &               &     24&.7(12)         \\
$h_3 \times 10^{12}$     &        &               &        &               &      3&.00(19)        \\
$L_{K} \times 10^{12}$   &        &               &        &               &  $-$64&.7(15)         \\
$L_{KKJ} \times 10^{12}$ &   $-$20&.51(15)        &   $-$33&.57(15)        &     18&.63(31)        \\
$L_{JK} \times 10^{12}$  &       0&.283(13)       &        &               &   $-$1&.257(13)       \\
$L_{JJK} \times 10^{15}$ &        &               &        &               &   $-$9&.4(12)         \\
$l_{1} \times 10^{15}$   &        &               &        &               &   $-$0&.718(82)       \\
$l_{2} \times 10^{15}$   &        &               &        &               &   $-$0&.327(71)       \\
$P_{KKJ} \times 10^{15}$ &        &               &        &               &   $-$0&.353(93)       \\
$\chi _{aa}$             &    $-$2&.726\,7(16)    &    $-$3&.643\,7(22)    &   $-$0&.041\,3(21)    \\
$\chi _{bb}$             &       0&.675\,3(17)    &       1&.987\,7(21)    &   $-$1&.935\,1(18)    \\
$\chi _{cc}$             &       2&.051\,4(20)$^c$&       1&.656\,0(19)$^c$&      1&.976\,4(16)$^c$\\
\hline
\end{tabular}\\[2pt]
\end{center}
$^a$ Watson's $S$ reduction has been used in the representation $I^r$. 
Numbers in parentheses are one standard deviations in units of the least 
significant figures.\\
$^b$ Estimated value, see Sect.~\ref{lab-discussion}.\\
$^c$ Derived value.\\
\end{table*}


The rotational spectra of the $AA$, $GA$, and $AG$ conformers of 
$n$-butyl cyanide were predicted and fit employing the {\scriptsize SPCAT} 
and {\scriptsize SPFIT} programs by \citet{spfit_1991}. The $^{14}$N 
hyperfine structure was not resolved in any of the present measurements 
because of the rather high quantum numbers and the larger line widths 
caused by using a free space cell rather than a molecular beam.

Our present measurements were started near 90~GHz. As shown in 
Fig.~\ref{lab-spectrum}, the spectrum is very dense, and lines 
belonging to low-$K_a$ transitions were displaced from the initial 
predictions by a few megahertz. The large number of lines recorded, 
the high line density, the presence of lines from three identified 
conformers, and very many unidentified lines made the analysis 
time-consuming and delicate particularly in the initial stages 
before improved predictions were available. 
Nevertheless, groups of closely spaced lines could be identified 
unambiguously because the predicted pattern was quite close to the 
observed one in terms of both frequencies and relative intensities. 
Improved predictions permitted us to assign transitions with higher 
$K_a$ even if they were isolated. Only transitions were used in the 
fit for which overlap by unassignable lines or lines from other 
conformers appeared to be negligible. In the 75.0$-$130.4~GHz region, 
$a$-type $R$-branch transitions were used in the fit for all conformers. 
Many $b$-type transitions were included for the $GA$ conformer, 
as can be expected from the large dipole moment component. 
In the case of the $AA$ conformer, the smaller $b$-dipole moment 
component initially permitted only $K_a = 1 \longleftrightarrow 0$ 
transitions to be assigned unambiguously in this frequency region. 
No $c$-type transitions were assignable unambiguously to the $AG$ 
conformer throughout all frequency regions studied in the course of 
the present investigation as the dipole moment component is even smaller 
and the lines of this conformer were weaker because of the much lower 
abundance than those of the $GA$ and $AA$ conformers.

The $J$ quantum number range was extended by additional measurements 
between 200.0~GHz and 222.3~GHz. It was straightforward to assign 
$a$-type transitions for all conformers as well as $b$-type transitions 
for the $GA$ conformer. In addition, a fairly tight $K_a = 8 - 7$ 
$Q$-branch was predicted for the $AA$ conformer in this frequency region 
whose relative intensity was quite favorable with respect to the lowest 
$K_a$ $a$-type $R$-branch transitions. Several transitions of that 
$Q$-branch could be identified unambiguously and permitted other 
$b$-type transitions to be identified not only in this frequency 
region, but also in the 75.0$-$130.4~GHz region.

Finally, spectral recordings obtained near 300~GHz and 350~GHz were 
analyzed. Large sections of the $K_a = 13 - 12$ $Q$-branch of the $AA$ 
conformer appeared to be overlapped negligibly by other features. 
In addition, some $c$-type transitions of the $AG$ conformer appeared 
to be sufficiently close to the predictions and have about 
the right intensities, but their number are insufficiently large 
for us to make unambiguous assignments. 
Altogether, more than 2200 transitions each have been recorded in 
the present investigation for the $AA$ and $GA$ conformers, while almost 
1400 transitions have been recorded for the higher energy $AG$ conformer. 
The number of distinct frequencies is smaller by about a factor of two, 
mainly because of the unresolved asymmetry splitting. The highest $J$ 
values of the transitions included in the fits are around 120 for the 
$GA$ and $AG$ conformers and even 136 for the $AA$ conformer because 
of the smaller value of $B + C$. The highest values of $K_a$ are 
higher than 40 for $GA$ and $AG$, and reach a slightly lower value of 
38 for the $AA$ conformer because of the larger $A$ rotational constant. 

Among the previous MB-FTMW measurements, the $2_{2,1} - 2_{1,2}$ 
hyperfine components of the $GA$ conformer were omitted from the fit 
because of the rather large and differing residuals. In addition, 
we excluded from the fit transitions with larger residuals for which 
no hyperfine splitting was resolved. These were the $8_{08} - 7_{07}$ 
transition of $AA$, the $7_{07} - 6_{06}$ transition of $AG$, and 
the $4_{23} - 3_{22}$ transition of $GA$.

Spectroscopic parameters were determined by employing Watson's $S$ 
reduction of the rotational Hamiltonian. The data from the previous 
MB-FTMW spectroscopic investigation~\citep{n-BuCN_rot_1997} were 
included in the fit with hyperfine splitting. Parameters were retained 
in the fit in general if they were determined with significance 
and if their inclusion contributed to the reduction in the rms error 
(the reduced $\chi ^2$). The resulting spectroscopic parameters 
are given in Table~\ref{spec-parameter}. The transition frequencies 
with their assignments, uncertainties, and residuals between observed 
frequency and that calculated from the final set of spectroscopic 
parameters are available in the supplementary material and are also 
available in the spectroscopy 
section\footnote{Internet address: http://www.astro.uni-koeln.de/cdms/daten} 
of the CDMS~\citep{CDMS_1,CDMS_2}.

The rms errors of the fits were slightly smaller than 1.0, meaning 
that the transition frequencies were reproduced on average within 
the experimental uncertainties. Moreover, partial rms errors for the 
MB-FTMW data and the present data around 100, 210, and above 300~GHz 
were mostly between 0.8 and 1.0 and did not exceed 1.1.

\subsection{Radioastronomical observations}
\label{obs-astro}

We used a complete line survey performed in the 3\,mm atmospheric 
window between 80 and 116~GHz toward the hot core region Sgr~B2(N). 
The observations were carried out in January 2004, September 2004, 
and January 2005 with the IRAM~30\,m telescope on Pico Veleta, Spain. 
Details about the observational setup and the data reduction are given 
in \citet{det_AAN}. An rms noise level of 15$-$20~mK on the 
$T^\star_{\mathrm{a}}$ scale was achieved below 100~GHz, 20$-$30~mK 
between 100~GHz and 114.5~GHz, and about 50~mK between 114.5~GHz and 
116~GHz. 

This survey aims to investigate the molecular complexity of this 
prolific source and characterize its molecular content. Overall, 
we detected about 3700 lines above $3\sigma$ over the whole 3\,mm band. 
These numbers correspond to an average line density of about 100 features 
per gigahertz. Given this high line density, the assignment of 
a line to a given  molecule can be trusted only if all lines emitted 
by this molecule in our frequency coverage are detected with the correct 
intensity predicted by a model of its excitation \textit{and} if none 
of the  predicted lines are missing in the observed spectrum. 
It is possible to search for new species once a sufficient number 
of lines emitted by known molecules have been identified, including 
vibrationally and torsionally excited states. The XCLASS 
software\footnote{See http://www.astro.uni-koeln.de/projects/schilke/XCLASS.}
is used to model the emission of all known molecules in the local 
thermodynamical equilibrium approximation (LTE for short), which refers 
here only to the rotational temperature from which the vibrational or 
conformational temperatures may differ. 
More details about this analysis are given in \citet{det_AAN}. About 
50 different molecules have been identified in Sgr~B2(N) thus far, 
and for several of them emission or absorption features due to 
minor isotopologs (about 60) or excited vibrational states (about 50) 
have also been identified (Belloche et al. in prep.). However, 
up to as many as 40\,\% of the significant lines remain unassigned, 
and some of them are rather strong. The article reporting the 
results of the full survey will be submitted later this year 
(Belloche et al., in prep.), and the data will become public after 
acceptance of the article.


\begin{figure}
\includegraphics[angle=-90,width=9cm]{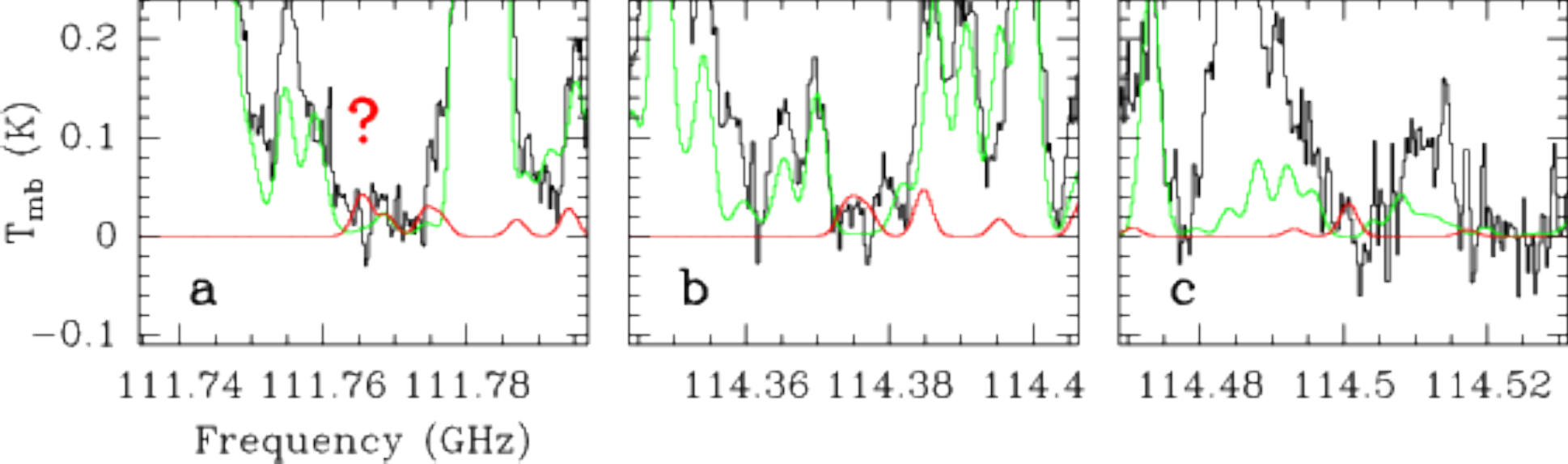}
  \caption{Sections of the 3\,mm line survey of Sgr~B2(N) showing the 
  LTE modeling of the astronomical observations. In each panel, the 
  black line represents the observed spectrum and the green line our 
  current model of all firmly assigned species, i.\,e. without 
  $n$-C$_4$H$_9$CN. The red line shows the model corresponding to the 
  parameters listed in Table~\ref{astro-modeling} for the $AA$ conformer 
  of $n$-butyl cyanide.}
\label{astro_spectra}
\end{figure}


\begin{figure}
\includegraphics[angle=-90,width=9cm]{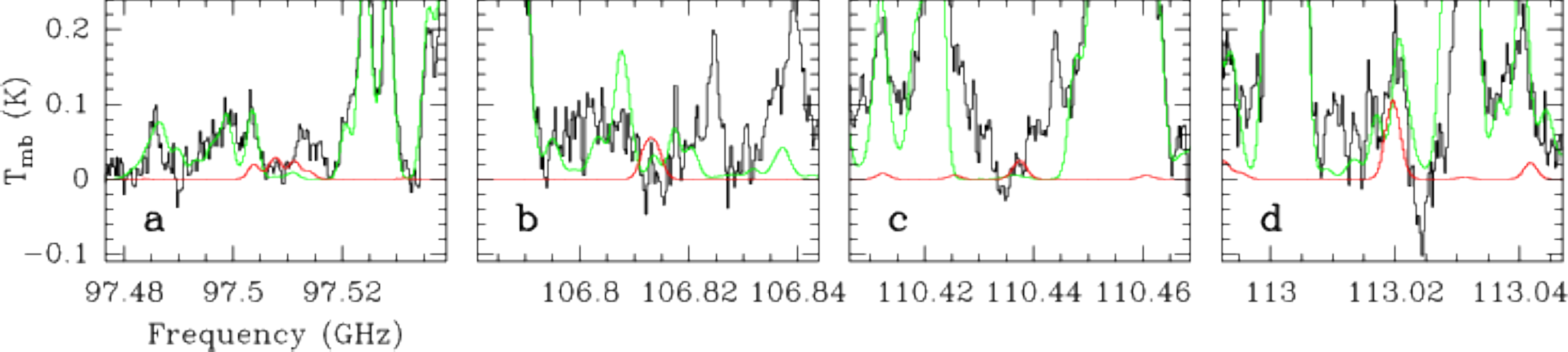}
  \caption{Same as Fig.~\ref{astro_spectra} but for the slightly 
   higher lying $GA$ conformer of $n$-butyl cyanide.}
\label{astro_spectra2}
\end{figure}



\begin{table}
\begin{center}
\caption{Parameters of our best-fit LTE models of alkyl cyanides with 
upper limits for $n$-butyl cyanide, which was not unambiguously detected.}
\label{astro-modeling}
\renewcommand{\arraystretch}{1.10}
\begin{tabular}[t]{lccrcr@{}l}
\hline \hline
\multicolumn{1}{c}{Molecule$^a$} & Size$^b$ & $T_{\rm{rot}}$$^c$ & \multicolumn{1}{c}{$N^d$}    
& $\Delta V^e$ & \multicolumn{2}{c}{$V_{\rm{off}}$$^f$} \\
                     & ('')     & (K)                & \multicolumn{1}{c}{(cm$^{-2}$)} 
& (km~s$^{-1}$)      & \multicolumn{2}{c}{(km~s$^{-1}$)} \\
 \multicolumn{1}{c}{(1)} & (2)  & (3)                & \multicolumn{1}{c}{(4)}      
& (5)          & \multicolumn{2}{c}{(6)}                \\
\hline
CH$_3$CN             & 2.7      & 200 & $2.0 \times 10^{18}$    & 6.5 &  $-$1&.0 \\
                     & 2.7      & 200 & $8.0 \times 10^{17}$    & 6.5 &     9&.0 \\
                     & 2.7      & 200 & $1.0 \times 10^{17}$    & 8.0 & $-$11&.0 \\
C$_2$H$_5$CN         & 3.0      & 170 & $1.2 \times 10^{18}$    & 6.5 &  $-$1&.0 \\
                     & 2.3      & 170 & $1.4 \times 10^{18}$    & 6.5 &     9&.0 \\
                     & 1.7      & 170 & $9.0 \times 10^{17}$    & 8.0 & $-$11&.0 \\
$n$-C$_3$H$_7$CN     & 3.0      & 150 & $1.5 \times 10^{16}$    & 7.0 &  $-$1&.0 \\
                     & 3.0      & 150 & $6.6 \times 10^{15}$    & 7.0 &     9&.0 \\
$AA$-$n$-C$_4$H$_9$CN& 3.0      & 150 &$\le 3.0 \times 10^{15}$ & 7.0 &  $-$1&.0 \\
$GA$-$n$-C$_4$H$_9$CN& 3.0      & 150 &$\le 8.0 \times 10^{15}$ & 7.0 &  $-$1&.0 \\
\hline
\end{tabular}\\[2pt]
\end{center}
Notes: $^a$ Results for CH$_3$CN, C$_2$H$_5$CN, and $n$-C$_3$H$_7$CN were taken from 
\citet{det-PrCN_EtFo}. $^b$ Source diameter ($FWHM$). $^c$ Temperature. 
$^d$~Column density or upper limit. $^e$ Linewidth ($FWHM$). $^f$ Velocity offset 
with respect to the systemic velocity of Sgr~B2(N), $V_{\rm{lsr}} = 64$~km s$^{-1}$.\\
\end{table}


We searched for emission features of the $AA$ and $GA$ conformers of 
$n$-butyl cyanide in our molecular line survey because these are 
the two conformers lowest in energy. The $AG$ conformer was deemed 
to be too high in energy to be observable under LTE conditions. Both 
$AA$ and $GA$ conformers were assumed to be ground state conformers 
since the exact energy difference between both conformers is unknown. 
As a consequence, the derived column density of the higher state conformer, 
$GA$ according to present knowledge, or the upper limit thereof, 
is likely overestimated. The source size, rotational temperature, 
line width, and finally the velocity offset from the systemic velocity 
were assumed to be the same as those for the stronger hot-core component 
of $n$-propyl cyanide (see Table~\ref{astro-modeling}). The parameters 
of the lighter alkyl cyanides are only slightly different.

The partition function values at 300~K and 150~K for the vibrational 
ground states of the $AA$ conformer of $n$-butyl cyanide were calculated 
as 174758.2 and 61661.6. The corresponding values for the $GA$ conformer, 
191118.2 and 67405.2, were calculated neglecting the conformational 
degeneracy as well as the non-zero energy of the lowest rotational 
state of the $GA$ conformer. 
Vibrational contributions to the partition function were also neglected 
because no information is available on the low-lying vibrational states 
of $n$-butyl cyanide conformers. These contributions are probably 
not negligible at 150~K and may be very substantial at 300~K, where 
the latter is the default temperature for the catalog entries and 
corresponds roughly to the laboratory conditions, and the former is 
the probable temperature of $n$-butyl cyanide in Sgr~B2(N) (see also 
Table~\ref{astro-modeling}). Additional partition function values will 
be provided in the CDMS.

Most transitions of both conformers are blended with stronger lines 
of other known species. However, a few transitions are relatively 
free of contamination and could in principle permit a detection. 
These are displayed in Figs.~\ref{astro_spectra} and \ref{astro_spectra2}. 
Two transitions of the $AA$ conformer coincide with lines that remain 
unidentified (Figs.~\ref{astro_spectra}b and c), but the prediction 
for another transition may be inconsistent with the observed spectrum 
depending on the exact position of the baseline, which is uncertain 
(Fig.~\ref{astro_spectra}a). As far as the $GA$ conformer is concerned, 
two transitions coincide with still unidentified lines 
(Figs.~\ref{astro_spectra2}a and c), one is partly blended 
with the weaker velocity component of a vibrationally excited state 
of C$_2$H$_5$CN (Fig.~\ref{astro_spectra2}b), and one with vibrationally 
excited dimethyl ether (Fig.~\ref{astro_spectra2}d). 

Overall, the number of potentially detected emission lines of either 
$n$-butyl cyanide conformer is too small to warrant even a tentative 
detection. As a result, we consider the models displayed in 
Figs.~\ref{astro_spectra} and \ref{astro_spectra2} as upper limits 
to the emission of $n$-butyl cyanide. Therefore, we derive upper limits 
of $3.0 \times 10^{15}$~cm$^{-2}$ and $8.0 \times 10^{15}$~cm$^{-2}$ 
for the column densities of the $AA$ and $GA$ conformers of 
$n$-butyl cyanide, respectively (see Table~\ref{astro-modeling}).

\section{Discussion}

\subsection{Laboratory spectroscopy}
\label{lab-discussion}

The present investigations permitted a full set of quartic centrifugal 
distortion parameters to be determined for the first time for each of 
the three $n$-butyl cyanide conformers previously characterized by 
MB-FTMW spectroscopy. In addition, several sextic distortion terms, 
up to a complete set for $GA$, along with some parameters of 
even higher order have been obtained with significance. 
Accurate predictions of the ground-state rotational spectrum are 
possible for strong to moderately weak transitions throughout 
the millimeter wave region and well into the submillimeter region 
for the two lower-energy conformers $AA$ and $GA$. In the case of 
the higher-energy $AG$ conformer, which was less well-characterized, 
predictions of $a$-type transitions should, nevertheless, be reliable 
up to possibly 500~GHz, but the $c$-type transitions are about 
as strong or stronger than the $a$-types beyond 300~GHz. However, they 
are difficult to assign because of the fairly large uncertainties 
for transitions with $K_a$ higher than 5, and they are still weaker 
than the transitions of the two more abundant conformers.

Since the two conformers, for which no rotational transitions have been 
identified thus far, are even higher in energy, it may be difficult to 
identify them in the recorded spectra. The best prospects should be at 
lower frequencies because of the fewer as well as narrower lines. 

The increase in $D_K$ and the decrease in both $D_J$ and $H_J$ 
from $AA$ to $AG$ and then $GA$ reflect the increase in $A$ and 
the decrease in $B + C$. A clear trend is also present for the 
$H_K$ values of the $AA$ and $GA$ conformers, permitting a value 
of 2.0~Hz to be estimated for the $AG$ conformer. The value may be 
lower by at least 0.5~Hz or higher by at least 1.0~Hz.
No clear trend is discernible for $D_{JK}$ as well as 
the higher-order diagonal distortion terms. Since $GA$ is much 
farther away from the prolate limit than both $AA$ and $AG$, 
off-diagonal distortion parameters, such as $d_1$, $d_2$, etc., 
play a much bigger role for this conformer than for the other two.

While almost all of the spectroscopic parameters of the three 
conformers have been firmly determined with small relative 
uncertainties, some parameters of the $GA$ conformer were just 
barely determined, most notably $l_2$ and $P_{KKJ}$. 
However, the values of both parameters appear to be reasonable. 
The decline in magnitude from $d_1$ to $d_2$ is much more than 
a factor of 10, while it is about a factor of 10 from $h_1$ to 
$h_2$ and from $h_2$ to $h_3$. Hence, it is conceivable that 
$l_1$ and $l_2$ differ by a factor only slightly larger than 2. 
Similarly, there is a factor of around 10000 from the magnitude 
of $D_{JK}$ to that of $H_{KJ}$ to $L_{KKJ}$, and finally to 
that of $P_{KKJ}$. It should be added that the correlation 
coefficients between $l_2$ or $P_{KKJ}$ with any of the other 
spectroscopic parameters are rather small in magnitude, except 
for those between $l_2$ and $h_2$ and between $P_{KKJ}$ and 
$L_{KKJ}$, which are fairly large, as expected. As a consequence, 
omission of $l_2$ or $P_{KKJ}$ affected only the corresponding 
lower order parameters, in particular $h_2$ or $L_{KKJ}$, 
respectively. However, these changes were deemed to be too large 
with respect to their uncertainties, and $l_2$ and $P_{KKJ}$ 
were retained in the fit. It is advisable to view their values 
with caution.

Infrared spectroscopy and low-level theoretical 
calculations~\citep{n-BuCN_IR_1989,n-BuCN_ai_2002} predict the 
$AA$ conformer to be the lowest in energy with the $GA$ conformer 
being only slightly higher. The relative energies of the conformers 
could also be determined from relative intensity measurements in the 
rotational spectra. In addition to at least reasonable knowledge of 
the rotational and vibrational contributions to the partition function, 
one would need to know the dipole moment components accurately. 
In the case of monocyanides of alkanes or alkenes, the dipole vector 
is close to being aligned with the CN bond, as in the examples of 
either vinyl and ethyl cyanide \citep{VyCN_EtCN_dip_2011} or 
$iso$-propyl cyanide \citep{i-PrCN_rot_2011}, and the total dipole 
moments are almost the same. The estimates of the dipole moment 
components for the $n$-butyl cyanide conformers, obtained by 
rotating the CH$_3$CN dipole vector such that it is parallel to 
the CN bond of the respective conformer \citep{n-BuCN_rot_1997}, 
are hence probably quite good. Small deviations from the alignment 
with the CN bond, however, as well as uncertainties about the 
structure may have non-negligible effects on the values of the 
dipole moment components. Therefore, it is desirable to carry out 
Stark effect measurements on the $n$-butyl cyanide conformers, 
which we hope to be able to carry out soon.

We note that higher level (MP2) calculations overestimate the stability 
of the CN group in the $gauche$ conformation compared to the $anti$ 
conformation. Unsurprisingly, a similar situation was encountered for 
MP2 calculations on $n$-propyl cyanide by \citet{n-PrCN_MP2_ED_2000}, 
who also describe a gas-phase electron-diffraction study of this 
molecule carried out at room temperature. The experimentally determined 
conformational composition, a $gauche$ to $anti$ ratio of about $3 : 1$, 
was very similar to the one calculated theoretically. However, these 
results were recognized to be at variance with results from a microwave 
study~\citep{n-PrCN_rot_egy_1988} in which relative intensities were 
measured at room temperature and 233~K. The $anti$ conformer was 
determined to be $1.1 \pm 0.3$~kJ/mol lower in energy than the 
$gauche$ conformer. A detection of $n$-propyl cyanide in Sgr~B2(N) 
by \citet{det-PrCN_EtFo} was restricted to lines of the $anti$ 
conformer, since transitions of the $gauche$ conformer were too weak 
at a rotational temperature of 150~K to be detected unambiguously. 
It was concluded that the $anti$ conformer may be even lower in 
energy than the $gauche$ conformer or that the molecules are not in LTE.

The energetics of the $n$-butyl cyanide conformers suggest that 
a detection in space may only be possible for the $AA$ conformer, 
that the $GA$ conformer has some chance of being detected, but that 
the $AG$ conformer, and even more so those not yet characterized 
by rotational spectroscopy, are likely too high in energy.

\subsection{Radioastronomical observations}
\label{astro-discussion}

Since $n$-propyl cyanide has been detected thus far only in the 
Sgr~B2(N) hot core, this source is presently the only viable source 
to search for $n$-butyl cyanide in space. Column density ratios of 
$108 : 80 : 1$ were derived for methyl, ethyl, and $n$-propyl 
cyanide~\citep{det-PrCN_EtFo}. On the basis of the two heavier 
molecules, one would expect a column density drop of about two orders 
of magnitude for $n$-butyl cyanide with respect to $n$-propyl 
cyanide. However, since methyl and ethyl cyanide have almost equal 
column densities, the column density of $n$-butyl cyanide may also 
be quite similar to that of $n$-propyl cyanide. The actual determination 
of the column density of $n$-butyl cyanide in Sgr~B2(N) will thus 
provide valuable clues to the formation of complex molecules in space.

The upper limits to the column density of the $AA$ and $GA$ 
conformers of $n$-butyl cyanide are lower by a factor of 5 and 
about 2, respectively, than the column density of $n$-propyl cyanide 
(see Table~\ref{astro-modeling}). The much lower, and thus more 
meaningful, upper limit to the $AA$ conformer than the $GA$ conformer 
can be explained by the smaller value of $B + C$ of 2598~MHz versus 
3343~MHz and the larger value of $\mu _a \approx 3.4$~D for $AA$ 
versus $\mu _a \approx 2.3$~D and $\mu _b \approx 3.1$~D for 
$GA$ \citep[][, see also Subsect.~\ref{obs}]{n-BuCN_rot_1997}.
Since current knowledge indicates that the $GA$ conformer is slightly 
higher in energy than the $AA$ conformer, the column density of the 
$GA$ conformer should be significantly less than twice that of the 
$AA$ conformer at rotational temperatures significantly below room 
temperature\footnote{All $n$-butyl cyanide conformers other than $AA$ 
are conformationally doubly degenerate, whereas $AA$ is not degenerate 
(see Sect.~\ref{intro}).}. If an energy difference of 0.54~kJ/mol 
(65~K) is assumed \citep{n-BuCN_IR_1989}, which is compatible with 
the abundances derived from IR spectroscopy, the column density ratio 
between the $AA$ and $GA$ conformers should be about $1.0 : 1.2$. 
On the basis of the upper limit of the $AA$ conformer, an upper 
limit to the total $n$-butyl cyanide column density of 
$6.6 \times 10^{15}$~cm$^{-2}$ would be derived for the main 
component of Sgr~B2(N). This is more than a factor of two lower 
than $n$-propyl cyanide. Hence, the possibility of very similar 
column densities of $n$-propyl cyanide and $n$-butyl cyanide, 
as mentioned above, may be unlikely. Decreasing the line confusion 
through interferometric observations should permit us to lower 
the column-density upper limit of $n$-butyl cyanide or could even 
lead to its detection. We hope to have observational results 
from ALMA soon.

Concerning radioastronomical searches for $n$-butyl cyanide more 
generally, the molecules aminoacetonitrile~\citep{det_AAN}, 
$n$-propyl cyanide, and ethyl formate~\citep{det-PrCN_EtFo} were 
detected most clearly at 3~mm because overlap with frequently 
stronger lines of lighter or more abundant molecules was not as 
significant in this frequency region as at either 2 or 1.3~mm. 
The frequencies are lower than or just reach the Boltzmann peaks 
of the $a$-type transitions at the derived rotational temperatures 
of 100~K for aminoacetonitrile ($\sim$230~GHz) and ethyl formate 
($\sim$190~GHz) or 150~K for $n$-propyl cyanide ($\sim$200~GHz). 
With the Boltzmann peak for the $a$-type transitions of the 
$AA$ conformer of $n$-butyl cyanide being at $\sim$150~GHz 
at 150~K, the molecule is probably most reliably searched for 
with a single-dish telecope at wavelengths longer than 3~mm. 
Interferometric observations, e.\,g. with the Plateau de Bure 
Interferometer (PdBI, to be upgraded to NOEMA), the Expanded Very 
Large Array (EVLA), or the Atacama Large Millimeter Array (ALMA), 
will alleviate the line confusion problem somewhat, which may make 
searches for the $AA$ conformer in space promising at 3~mm 
or maybe even shorter wavelengths. At any rate, the current 
laboratory measurements together with the previous ones as well as 
interpolations should cover all frequencies suitable for the search 
for $n$-butyl cyanide in space; moreover, extrapolation to 
higher frequencies is reasonable to some extent.

\section{Conclusion}
\label{conclusion}

The rotational spectra of three low-lying conformers of $n$-butyl cyanide 
have been studied extensively in the millimeter and lower submillimeter 
regions, providing the means to search for this molecule in space. 
Inspection of our sensitive 3~mm line survey toward Sgr~B2(N), carried 
out with the IRAM 30~m telescope, yielded an upper limit to the column 
density of the lowest energy $AA$ conformer that is considerably lower 
than the column density found recently for the shorter $n$-propyl cyanide 
\citep{det-PrCN_EtFo}. Observations with ALMA, other telescope arrays, 
or single-dish telescopes at lower frequencies should alleviate the 
line confusion, leading to a considerable lowering of the upper limit 
to or the actual detection of the molecule, either of which will 
provide important clues about the molecular complexity in space. 
Observational constraints on the column density of $iso$-propyl cyanide, 
a molecule that has also been studied recently in our laboratory 
\citep{i-PrCN_rot_2011}, will also be interesting as it is the 
smallest branched cyanide. The detection of this molecule 
or sufficiently low upper limits to its column densities 
will provide important information on the importance of 
branched molecules with respect to their unbranched isomers. 
Several other complex molecules may also be detectable in this 
prolific source. It is thus likely that observations of Sgr~B2(N) 
with ALMA around 3~mm will provide deeper insight into astrochemistry.


\begin{acknowledgements}
We thank Prof. R.~K. Bohn for providing the $n$-butyl cyanide 
transition frequencies with observed $^{14}$N quadrupole splitting. 
The present investigations have been supported by the Deutsche 
Forschungsgemeinschaft (DFG) in the framework of the collaborative 
research grant SFB~956, project B3. 
H.S.P.M. is very grateful to the Bundesministerium f\"ur Bildung und 
Forschung (BMBF) for financial support through project FKZ 50OF0901 
(ICC HIFI \textit{Herschel}) aimed at maintaining the Cologne Database 
for Molecular Spectroscopy, CDMS. This support has been administered 
by the Deutsches Zentrum f\"ur Luft- und Raumfahrt (DLR). 
A.W. and M.N. thank the French National Program PCMI (CNRS/INSU) 
for funding.
\end{acknowledgements}



\end{document}